\documentclass[preprint, aps, showpacs, amsmath, amssymb, nofootinbib, footnotes, superscriptaddress]{revtex4}
\usepackage{graphicx}
\usepackage{bm}
\usepackage{color}

\begin{document}

\title{Manifestation of a singlet heavy up-type quark in the branching ratios
of rare decays $K \rightarrow \pi \nu \bar{\nu}$, $B \rightarrow
\pi \nu \bar{\nu}$ and $B \rightarrow K \nu \bar{\nu}$}
\author{P.\ N. Kopnin}
\email{kopnin@itep.ru} \affiliation{Institute for Theoretical and
Experimental Physics, Moscow, Russia} \affiliation{Moscow
Institute of Physics and Technology (State University)}
\author{M.\ I. Vysotsky}
\email{vysotsky@itep.ru} \affiliation{Institute for Theoretical
and Experimental Physics, Moscow, Russia}

\begin{abstract}
We investigate the implications of the model with a
$SU(2)$--singlet up-type quark, heavy enough not to be produced at
the LHC, namely, the contribution of the new quark to the
branching ratios of the $K \rightarrow \pi \nu \bar{\nu}$, $B
\rightarrow \pi \nu \bar{\nu}$ and $B \rightarrow K \nu \bar{\nu}$
decays. We show that the deviation from the Standard Model can be
up to $10\%$ in the case of a $5\rm\ TeV$ quark. Precise
measurements of these branching ratios at the future experiments
will allow to observe the contributions of the new quark or to
impose stronger constraints on its mass.
\end{abstract}

\maketitle

\newpage

\section{Introduction.}
When we discuss the corrections to the observables in flavor
physics due to various types of New Physics, the processes that we
investigate are determined by loop diagrams of two types:
box--diagrams and the so-called penguin diagrams. In
\cite{vysotsky}, \cite{me} the contribution of a singlet heavy
up-type quark to the observables determined by the box--diagrams (the
mass differences of the neutral B--mesons $\Delta m_{B_d}$,
$\Delta m_{B_s}$ and CP--violation parameter in the oscillations
of K--mesons $\varepsilon_K$) is discussed. \footnote{In the paper
\cite{nir} the properties of the $SU(2)$--singlet down-type quarks
within the framework of the Minimal Flavor Violation are
discussed. In \cite{deshpande} the tree-level contributions to
different observables of a down-type singlet quark are
investigated. A model with a singlet up-type quark was studied in
\cite{alwall} in which the additional quark is comparatively light
(just above the reach of the Tevatron) and mixes strongly with the
$t$--quark. In \cite{aguilar} a model with a relatively light
(less than $600\ \rm GeV$) singlet up-type
quark is discussed. The formulae of \cite{aguilar} lead to a wrong
asymptotic behavior of the branching ratios under consideration
in the case of a heavy quark.} The rare decays $K \rightarrow \pi \nu \bar{\nu}$, $B
\rightarrow \pi \nu \bar{\nu}$ and $B \rightarrow K \nu \bar{\nu}$
are the processes determined in the Standard Model (SM) by penguin
diagrams.

The possibility of the presence of new particles inside the loops
makes the observables determined by loop diagrams quite useful
when obtaining experimental bounds on the parameters of the models
of New Physics. But the most important point here is the
following. The formulae for the observables $\Delta m_{B_d}$,
$\Delta m_{B_s}$ and $\varepsilon_K$ involve the pseudoscalar
decay constants $f(B_d,\ B_s)$ and bag parameters $B(K,\ B_d,\
B_s)$ defined (in the case of the $B_d$--meson) as follows:
\begin{eqnarray}
&&\langle 0| [\bar{b}\gamma_\mu(1+\gamma^5)d] |B_d\rangle =
if_{B_d}p^{B_d}_{\mu}, \nonumber\\
 &&\langle \bar{B_d}|
[\bar{b}_L\gamma_\mu d_L][\bar{b}_L\gamma_\mu d_L] |B_d
\rangle=\frac{8}{3} B_{B_d}  \langle \bar{B_d}|
[\bar{b}_L\gamma_\mu d_L]|0\rangle \langle 0|[\bar{b}_L\gamma_\mu
d_L] |B_d \rangle. \label{f,B}
\end{eqnarray}
The accuracy with which they are known is quite poor. For example,
the evaluation of $f_B^2$ and $B(B_d,\ B_s)$ based on QCD lattice
calculations has a $10\%$ accuracy. Theoretical uncertainties of
this kind make it impossible to detect the contributions of the
New Physics to these observables if they are not well above
$10\%$.

The theoretical expressions for the decay widths $\Gamma(K^+
\rightarrow \pi^+ \nu \bar{\nu})$, $\Gamma(K_L \rightarrow \pi^0
\nu \bar{\nu})$, $\Gamma(B_u \rightarrow \pi^+ \nu \bar{\nu})$ and
$\Gamma(B_d \rightarrow \pi^0 \nu \bar{\nu})$ include matrix
elements $\langle \pi^+| \bar{s}_L\gamma_{\mu}d_L | K^+ \rangle$,
$\langle \pi^0| \bar{s}_L\gamma_{\mu}d_L | K^0 \rangle$, $\langle
\pi^+| \bar{b}_L\gamma_{\mu}d_L | B_u \rangle$ and $\langle \pi^0|
\bar{b}_L\gamma_{\mu}d_L | B_d \rangle$. Fortunately, these matrix
elements are equal to the matrix elements $\langle \pi^0|
\bar{s}_L\gamma_{\mu}u_L | K^+ \rangle$, $\langle \pi^-|
\bar{s}_L\gamma_{\mu}u_L | K^0 \rangle$, $\langle \pi^0|
\bar{b}_L\gamma_{\mu}u_L | B_u \rangle$ and $\langle \pi^-|
\bar{b}_L\gamma_{\mu}u_L | B_d \rangle$ respectively with the
accuracy of the isospin $SU(2)$ symmetry violation, which is
approximately $\dfrac{m_u-m_d}{\Lambda_{QCD}} \approx 1\%$. The
latter can be extracted from the data on the $K^+ \rightarrow
\pi^0 \nu e^+$, $K^0 \rightarrow \pi^-\nu e^+$, $B_u \rightarrow
\pi^0 \nu e^+$ and $B_d \rightarrow \pi^- \nu e^+$ decay widths.
(For the $B \rightarrow K \nu \bar{\nu}$ width the corresponding
accuracy is worse --- of order of the $SU(3)$ symmetry violation
$\approx 20\%$.) For this reason the branching ratios $Br(K^+
\rightarrow \pi^+ \nu \bar{\nu})$, $Br(K_L \rightarrow \pi^0 \nu
\bar{\nu})$, $Br(B_u \rightarrow \pi^+ \nu \bar{\nu})$ and $Br(B_d
\rightarrow \pi^0 \nu \bar{\nu})$ are well calculable, can serve
as indicators of New Physics and help to establish its parameters.




\section{Neutral and charged currents in the extended model.}

We are discussing a model with New Physics proposed in
\cite{vysotsky}. For its detailed description see \cite{vysotsky}
and  \cite{me}. Now we will present the formulae necessary for the
calculations using the notations of \cite{me}.

Our model Lagrangian is the following \footnote{${\cal L}_{SM}$ is
the SM Lagrangian with a zero Yukawa coupling of the Higgs to the
$t$--quark.} (\cite{vysotsky}, \cite{me}):
\begin{eqnarray}
{\cal L} &=& {\cal L}_{SM} +
\bar{Q'}\left(i\gamma^{\mu}D_{\mu}-M\right)Q' +
\left[\mu_R\bar{Q'}_Lt'_R+\frac{\mu_L}{\eta/\sqrt{2}}H_c^+\bar{Q'}_R
\dbinom{t'}{b^V}_L + c.c.\right]. \label{lagrangian}\end{eqnarray}
Here
$D_{\mu}=\partial_{\mu}-i\frac{2}{3}g'B_{\mu}-ig_sG^a_{\mu}\frac{\lambda^a}{2}$.
$ t'_L = U_{t't''}t''_L+U_{t'c'}c'_L+U_{t'u'}u'_L$, where $t''_L,
c'_L, u'_L$ are the fields of the SM in the so-called flavor
basis, $U$ is the matrix that rotates them into the mass
eigenstates $c_L, u_L$ and $t'_L$ ($t'$ would have a zero mass if
not for the mixing with the $Q$), and
\[
b^V_L = V_{tb}b_L+V_{ts}s_L+V_{td}d_L,
\]
where $b_L, s_L, d_L$ are the fields in the mass basis, $V$ is the
CKM (Cabibbo--Kobayashi--Maskawa) matrix. $H_c$ is the
$SU(2)$--conjugate of the Higgs isodoublet $H$.

To establish the connection between the $Q'$ and $t'$ fields and
the mass eigenstates one has to diagonalize the relevant bilinear
terms of the Lagrangian. The result is the following:

\begin{equation}
t' \equiv t'_R + t'_L= N_Rt_R - \frac{m_t}{\mu_L}N_LQ_R + N_Lt_L -
\frac{m_t}{\mu_R}N_RQ_L,
 \label{t'}\end{equation}

 \begin{equation}
Q' \equiv Q'_R + Q'_L= N_RQ_R + \frac{m_t}{\mu_L}N_Lt_R + N_LQ_L +
\frac{m_t}{\mu_R}N_Rt_L,
 \label{Q'}\end{equation}
where
  \begin{eqnarray}
&&N_L =
\left[1+\frac{\mu_L^2}{M^2}\left(1-\frac{m_t^2}{\mu_L^2}\right)^2\right]^{-1/2},\nonumber\\
&&N_R =
\left[1+\frac{\mu_R^2}{M^2}\left(1-\frac{m_t^2}{\mu_R^2}\right)^2\right]^{-1/2}.
\label{N}\end{eqnarray}

The masses of $Q$ and $t$ are the following:
\begin{equation}
m_Q = M + O\left(\frac{\mu^2}{M}\right),\ m_t =
\frac{\mu_L\mu_R}{m_Q},
 \label{masses}\end{equation}
and the whole mass of the $t$--quark originates from its mixing
with the heavy singlet quark $Q'$.

Substituting the expressions for the primed fields into the
Lagrangian one gets the explicit form of the charged current (CC)
interactions of the $t$-- and $Q$--quarks with the gauge ($W$) and
goldstone ($G$) bosons\footnote{$G$ is the charged unphysical
Higgs field that appears in the 't Hooft $R_{\xi}$--gauge}:
\begin{eqnarray}
{\cal L}_{CC} &=& \left(
\frac{g}{\sqrt{2}}\bar{b}^V_L\gamma_{\mu}W^{\mu}t_L +
\frac{m_t\sqrt{2}}{\eta}\bar{b}^V_LGt_R \right)N_L\nonumber\\
&-&\left( \frac{g}{\sqrt{2}}\bar{b}^V_L\gamma_{\mu}W^{\mu}Q_L +
\frac{m_Q\sqrt{2}}{\eta}\bar{b}^V_LGQ_R
\right)N_R\frac{\mu_L}{m_Q} + c.c., \label{QtWG}\end{eqnarray} ---
see \cite{me}.

\begin{figure}[tbh]
\centerline{\includegraphics[width=0.8\linewidth]{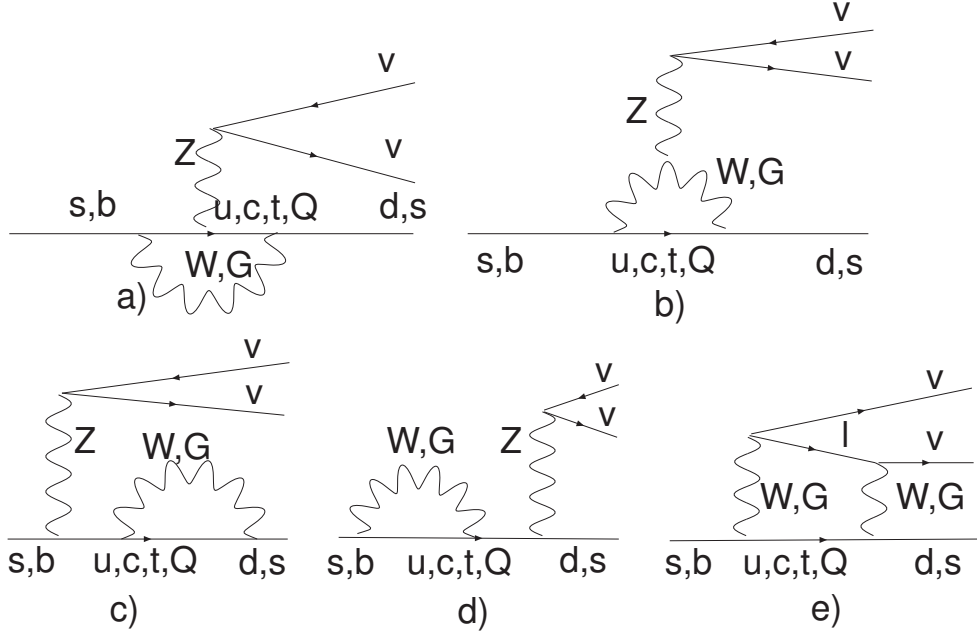}}
\caption{Diagrams contributing to the $s \rightarrow d \nu
\bar{\nu}$, $b \rightarrow d \nu \bar{\nu}$ and $b \rightarrow s \nu \bar{\nu}$
transitions in the extension of the SM. Diagrams with self-energy
insertions (c, d) have to be multiplied by $\dfrac{1}{2}$.}
\label{fullpenguinextsm}
\end{figure}

The penguin decays $K \rightarrow \pi \nu \bar{\nu}$, $B
\rightarrow \pi \nu \bar{\nu}$ and $B \rightarrow K \nu \bar{\nu}$
mentioned above are described by five diagrams (Fig.
\ref{fullpenguinextsm}).


In our model besides the $u$--, $c$-- and $t$--quarks we will have
to take into account the $Q$--quark exchange as well. The diagrams
include the interaction of the quarks with the $W$--boson and the
goldstone $G$. The first four of them also include the $Z$--boson
exchange between the fermionic currents. To calculate them we have
to establish the way in which the new particle $Q$ interacts with
$Z$.

The interaction of the $Q$--quark with $Z$ will arise from the
following terms of the Lagrangian\footnote{$\hat{{\cal Q}}$ stands
for the electric charge operator and it is not to be confused with
the quark $Q$.} :
\begin{eqnarray}
{\cal L}_{Q, t \leftrightarrow Z} &=& i\bar{g}Z_{\mu}\left[
\bar{t'}\gamma^{\mu}(\hat{T_3}-\hat{{\cal
Q}}\sin^2\theta_W)t'
-\frac{2}{3}\sin^2\theta_W\cdot\bar{Q'}\gamma^{\mu}Q'
\right]. \label{Q't'Z}\end{eqnarray} Taking into account
(\ref{t'}), (\ref{Q'}) and (\ref{N}) we obtain the following
expression for the $ttZ$, $QQZ$ and $tQZ$ neutral current (NC)
interactions in our model:
 \begin{eqnarray}
{\cal L}_{NC} &=& i\bar{g}Z_{\mu}\left[
\frac{1}{2}N_L^2\cdot\bar{t}_L\gamma^{\mu}t_L
-\frac{2}{3}\sin^2\theta_W\cdot\bar{t}\gamma^{\mu}t\right.
-\frac{2}{3}\sin^2\theta_W \cdot \bar{Q}\gamma^{\mu}Q +
\frac{1}{2}\frac{\mu_L^2}{m_Q^2}N_R^2 \cdot
\bar{Q}_L\gamma^{\mu}Q_L \nonumber\\
&-&\left. \frac{1}{2}\frac{\mu_L}{m_Q}N_RN_L \cdot \left(
\bar{t}_L\gamma^{\mu}Q_L
 + \bar{Q}_L\gamma^{\mu}t_L \right)\right].
\label{QtZ}\end{eqnarray} In our model the flavor changing $tQZ$
neutral current (FCNC) appears.

Just like in \cite{vysotsky}, \cite{me}, throughout the paper we
use the following numerical values:
\begin{eqnarray}
M=5\ \rm TeV,\ \mu_L= 500\ \rm GeV,\ \mu_R = 1.7\ \rm TeV,\ N_L =
0.996,\ N_R = 0.946.\label{numbers}
\end{eqnarray}



\section{The effective Lagrangian of the $K \rightarrow \pi \nu \bar{\nu}$
decay.}

In the SM both the $c$-- and the $t$--quarks contribute to the
penguin diagrams. The contribution of each of the up quarks is
proportional to $m_i^2\cdot V_{is}V^*_{id}$. Using the data from
\cite{pdg}, we find out that $\dfrac{m_t^2}{m_c^2} \approx 2\times
10^{4}$, $\dfrac{|V_{ts}V^*_{td}|}{|V_{cs}V^*_{cd}|} \approx
1.2\times 10^{-3}$, but the contribution of the $c$--quark is
numerically enhanced as compared to that of the $t$--quark and in
general we have to take into account both of them. Nevertheless,
for the reasons discussed in the following section in the case of
the decays $K_L \rightarrow \pi^0 \nu \bar{\nu}$, $B \rightarrow
\pi \nu \bar{\nu}$ and $B \rightarrow K \nu \bar{\nu}$ the
$t$--quark dominates.

The expression for the ${\cal L}_{eff}(s \rightarrow d \nu
\bar{\nu})$ in the SM was obtained in \cite{inamilim} (also
\cite{buras}):
\begin{eqnarray}
&&{\cal L}_{eff}^{SM}(s \rightarrow d \nu \bar{\nu}) = {\cal
L}_1(m_t) + {\cal L}'_1(m_c, m_l);\nonumber\\
&&{\cal L}_1(m_t) =\frac{G_F^2m_W^2}{4\pi^2}\
\bar{d}_L\gamma^{\mu}s_L \sum\limits_{l=e, \mu, \tau}
\bar{\nu}_L^{(l)}\gamma_{\mu}\nu_L^{(l)}
\cdot V^*_{td}V_{ts}\xi_t\ F(\xi_t)\ \eta_X; \nonumber\\
&&{\cal L}'_1(m_c, m_l) = \frac{G_F^2m_W^2}{4\pi^2}\
\bar{d}_L\gamma^{\mu}s_L \sum\limits_{l=e, \mu, \tau}
\bar{\nu}_L^{(l)}\gamma_{\mu}\nu_L^{(l)}\cdot V^*_{cd}V_{cs}\ X^l(m_l,
m_c),\label{L_sdnunu_SM}
\end{eqnarray} here $\xi_t \equiv \dfrac{m_t^2}{m_W^2}$, $m_t$ is the mass of the
top quark.
\begin{equation}
F(\xi) \equiv \left[ \frac{\xi+2}{\xi-1} +
\frac{3\xi-6}{(\xi-1)^2}\ \ln\xi \right] \label{Fxi}
\end{equation} numerically equals  $F(\xi_t) =
2.744$. The sum is over the neutrino flavors $l=e, \mu, \tau$ and
$X^l(m_l, m_c)$ accounts for the charm contribution. The factor
$\eta_X$ summarizes the QCD corrections, $\eta_X = 0,995 \approx
1$ \cite{buras}.

When calculating the diagrams (Fig. \ref{fullpenguinextsm}) we
will neglect the masses and momenta of the external quarks
compared to the momenta and masses of the particles inside the
loop. When the external momenta are neglected, the diagrams c) and
d) yield the same result.

Our model with a heavy singlet quark $Q$ modifies the result of
the SM in two ways: firstly, we have to take into account the
modification of the coupling of $t$--quark with $Z$--boson
(\ref{QtZ}) and with $W$-- and $G$--bosons (\ref{QtWG}), secondly,
we have to take into account the new particle $Q$.

The charged current interactions of the $Q$ and $t$ have the form
described in (\ref{QtWG}). The neutral currents (NC) in our model
(\ref{QtZ}) consist of four terms:
\begin{enumerate}
\item the current proportional to the charge of the particles
$\hat{{\cal Q}}$; \item the left $t$--quark current with the
coupling $\dfrac{1}{2} N_L^2$; \item the left $Q$--quark current
with the coupling $\dfrac{1}{2}\dfrac{\mu_L^2}{m_Q^2}N_R^2$; \item
the left FCNC ($QtZ$) with the coupling
$-\dfrac{1}{2}\dfrac{\mu_L}{m_Q}N_RN_L$.
\end{enumerate}

The first term --- proportional to $\hat{{\cal Q}}$ (the operator
of the electric charge) --- does not contribute to the penguin
amplitude. The crucial observation is that we are dealing here
with a conserved vector current. Since all the momenta of the
external particles ($s, d, Z$) are set to zero, we are dealing
with four diagrams (Fig. \ref{fullpenguinextsm} , a--d), the sum
of which is zero. The first two of them give the renormalization
of the vertex $\rm Z_1^{-1}-1$ and the last two
--- the renormalization of the fermion wave-function
$2\cdot\dfrac{1}{2}{\rm Z_2}-1$. The Ward identity in the abelian
case leads to the equality $\rm Z_1 = Z_2$. This implies that this
part of the decay amplitude is equal to its tree-level value, i.e.
zero.

\begin{figure}[tbh]
\centerline{\includegraphics[width=0.55\linewidth]{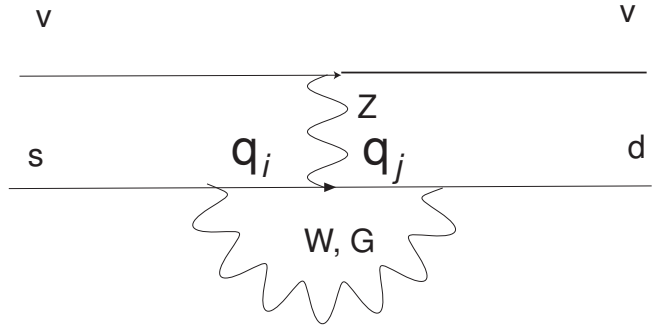}}
\caption{The diagram that defines the expression ${\cal
P}(m_i, m_j)$. The up-type quarks $q_i$ and $q_j$ have masses
$m_i$ and $m_j$ respectively.}
 \label{pdiag}
\end{figure}

The form of the CC interactions of the $t$--quark (\ref{QtWG})
indicates that all the SM diagrams have to be multiplied by
$N_L^2$. The second term of the NC implies that the diagram a)
with the $t$--quark in Fig. \ref{fullpenguinextsm} has an extra
$N_L^2$ factor amounting to a total factor $N_L^4$. In the same
way, the diagrams with $Q$--quark will have the factor
$\left(\dfrac{\mu_L}{m_Q}N_R\right)^2$, while for the diagram a)
with $Q$--quark the corresponding factor is
$\left(\dfrac{\mu_L}{m_Q}N_R\right)^4$. The same diagram with $Qt$
FCNC has the factor $\left(\dfrac{\mu_L}{m_Q}N_R N_L\right)^2$.
This suggests the following way of taking into account the effects
of New Physics:
\begin{enumerate}
\item to multiply the result of the SM by $N_L^2$ obtaining ${\cal
L}_1(m_t)\cdot N_L^2$; \item to calculate the diagrams with $Q$--quark
which do not include the $QtZ$ FCNC (they include the factor
$\left(\dfrac{\mu_L}{m_Q}N_R\right)^2$) obtaining ${\cal
L}_2(m_Q)\cdot \left(\dfrac{\mu_L}{m_Q}N_R\right)^2$, \item in order
to take into account the $QtZ$ FCNC to calculate diagram in Fig.
\ref{pdiag} --- ${\cal P}(m_i, m_j)$ (it coincides with the
diagram a) in Fig. \ref{fullpenguinextsm} when $m_i$ and $m_j$ are
equal to $m_t$ and $m_Q$).
\end{enumerate}

The expression for the effective Lagrangian ${\cal L}_{eff}^{NP}(s
\rightarrow d \nu \bar{\nu})$ which takes into account the
contribution of the New Physics obtained in this way is the
following:
\begin{eqnarray}
&&{\cal L}_{eff}^{NP}(s \rightarrow d \nu \bar{\nu}) = {\cal
L}_1(m_t)\ N_L^2 + {\cal
L}_2(m_Q)\left(\dfrac{\mu_L}{m_Q}N_R\right)^2
+ {\cal P}(m_t, m_t)\ (N_L^4-N_L^2) + {\cal P}(m_Q, m_Q) \nonumber\\
&\times&\left[\left(\dfrac{\mu_L}{m_Q}N_R\right)^4 -
\left(\dfrac{\mu_L}{m_Q}N_R\right)^2\right]
+ 2{\cal P}(m_t, m_Q)\left(\dfrac{\mu_L}{m_Q}N_R N_L\right)^2 = {\cal L}_1(m_t)N_L^2
+ {\cal L}_2(m_Q)\left(\dfrac{\mu_L}{m_Q}N_R\right)^2 \nonumber\\ &+&
\left(\dfrac{\mu_L}{m_Q}N_R N_L\right)^2\left( 2{\cal P}(m_t, m_Q)
- {\cal P}(m_t, m_t) - {\cal P}(m_Q, m_Q)
\right).\label{L_sdnunu_NP_def}
\end{eqnarray} Here we have used the identity $N_L^2 + \left(\dfrac{\mu_L}{m_Q}N_R\right)^2 =
1$ that can be easily established (\ref{N}).

Let us investigate the contribution of the $Q$--quark. Since
$\dfrac{m_Q^2}{m_W^2} \approx 4\times 10^3$, the $m_Q^2 \gg m_W^2$
limit is applicable while calculating the diagrams with the
$Q$--quark. In this limit we can neglect the $W$--boson exchanges
in the diagrams; because of the heaviness of the $Q$ the leading
contribution will come from the interaction with the unphysical
higgses $G$ (the interaction with them is $m_Q/m_W$ times stronger
than the interaction with the gauge bosons $W$, (\ref{QtWG})). It
is also important to note that taking only the $G$ exchanges into
account we will have to neglect the fifth diagram in Fig.
\ref{fullpenguinextsm} (i.e. the lepton box diagram), because its
contribution is proportional to the small ratio
$\dfrac{m_l^2}{\eta^2}$, where $m_l$ is the charged lepton mass.

\begin{figure}[tbh]
\centerline{\includegraphics[width=0.7\linewidth]{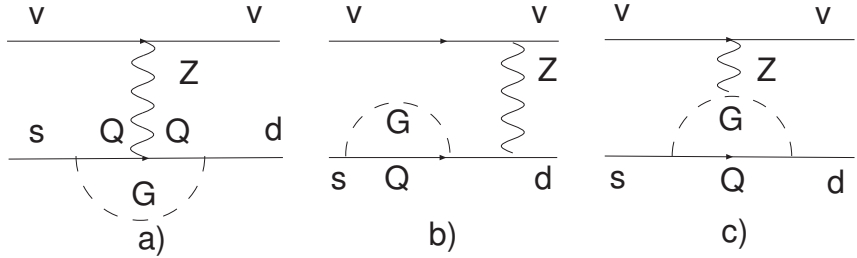}}
\caption{Diagrams contributing to the $s \rightarrow d \nu
\bar{\nu}$ transition with the $Q$--quark in the $m_Q^2 \gg m_W^2$
limit.} \label{extsmgoldstonesdiag}
\end{figure}

Thus, we come to the diagrams shown in Fig.
\ref{extsmgoldstonesdiag}.

The diagrams a -- c in Fig. \ref{extsmgoldstonesdiag} correspond
to the ${\cal L}_2(m_Q)$ part of the interaction of the $Q$ with
the $Z$--boson and yield the result:
\begin{eqnarray}
{\cal L}_2(m_Q) &=& \frac{G_F^2}{4\pi^2}\
\bar{d}_L\gamma^{\mu}s_L\ \sum\limits_l
\bar{\nu}_L^{(l)}\gamma_{\mu}\nu_L^{(l)}\cdot
V^*_{td}V_{ts}m_Q^2.\label{delta_L_diag}
\end{eqnarray}

The expression $2{\cal P}(m_t, m_Q) - {\cal P}(m_t, m_t) - {\cal
P}(m_Q, m_Q)$ equals:
\begin{eqnarray}
&&2{\cal P}(m_t, m_Q) - {\cal P}(m_t, m_t) - {\cal P}(m_Q, m_Q) =
Z_{\mu}\frac{G_F^2m_W^2}{4\pi^2}\ \bar{d}_L\gamma^{\mu}s_L
V^*_{td}V_{ts}\ \sum\limits_l
\bar{\nu}_L^{(l)}\gamma_{\mu}\nu_L^{(l)}\nonumber\\
&\times&\left[ \left( 2(\xi_t-1)\ln\xi_Q -
\frac{2\xi_t^2}{\xi_t-1}\ln\xi_t \right)\right. - \left.\left(
\xi_t - \frac{2}{\xi_t-1}\ln\xi_t \right) - \xi_Q
\right].\label{delta_L_nondiag}
\end{eqnarray}

The resulting expression for the effective Lagrangian is:
\begin{eqnarray}
&&{\cal L}_{eff}^{NP}(s \rightarrow d \nu \bar{\nu}) =
\frac{G_F^2}{4\pi^2}\ \bar{d}_L\gamma^{\mu}s_L \sum\limits_l
\bar{\nu}_L^{(l)}\gamma_{\mu}\nu_L^{(l)}
V^*_{td}V_{ts}m_t^2\nonumber\\
&\times& \left[F(\xi_t) N_L^4 + \frac{\mu_L^4}{m_t^2m_Q^2}
N_R^4\right. + \left.
\left(\frac{\xi_t-1}{\xi_t}\ln\frac{m_Q^2}{m_t^2} -
G(\xi_t)\right)\cdot
2\frac{\mu_L^2}{m_Q^2}\ N_R^2N_L^2\right] \nonumber\\
&=& {\cal L}_1(m_t)\left[N_L^4 + \frac{\mu_L^4}{m_t^2m_Q^2} N_R^4
F(\xi_t)^{-1}+ 2\frac{\mu_L^2}{m_Q^2}
\left(\frac{\xi_t-1}{\xi_t}\ln\frac{m_Q^2}{m_t^2} -
G(\xi_t)\right)N_R^2N_L^2 F(\xi_t)^{-1}\right]\label{L_sdnunu_NP}.
\end{eqnarray} Here the function $G(\xi)$ is defined as follows:
\begin{eqnarray}
G(\xi) = \frac{\xi+1}{\xi(\xi-1)}\ln\xi \label{Gxi}
\end{eqnarray} and numerically $G(\xi_t) = 0.513$. It is important
to note that in the final expression the term proportional to
$\left(\dfrac{\mu_L}{m_Q}N_R N_L\right)^2\cdot m_W^2\xi_Q$ arising
from the last term in (\ref{delta_L_nondiag}) almost completely
cancels the anomalously big contribution proportional to
$\left(\dfrac{\mu_L}{m_Q}N_R\right)^2\cdot m_Q^2$ from
(\ref{delta_L_diag}), giving the resulting term
$\left(\dfrac{\mu_L}{m_Q}N_R\right)^4\cdot m_Q^2 =
\dfrac{\mu_L^4}{m_t^2m_Q^2}N_R^4\ m_t^2$ in (\ref{L_sdnunu_NP}).




\section{Comparison with experimental data: prospects.}
The quantity $\delta_p$ that describes the deviation of the ${\cal
L}_{eff}^{NP}(s \rightarrow d \nu \bar{\nu})$ from the ${\cal
L}_{eff}^{SM}(s \rightarrow d \nu \bar{\nu})$ (i.e. ${\cal
L}_{eff}^{NP}(s \rightarrow d \nu \bar{\nu}) =
(1+\delta_p)\cdot{\cal L}_{eff}^{SM}(s \rightarrow d \nu
\bar{\nu})$) equals (\ref{L_sdnunu_SM}), (\ref{L_sdnunu_NP}):
\begin{eqnarray}
\delta_p &=& - \delta_c + N_L^4 -1 +
\frac{\mu_L^4}{m_t^2m_Q^2}\
N_R^4 F(\xi_t)^{-1}\nonumber\\
 &+& 2\frac{\mu_L^2}{m_Q^2}\
\left(\frac{\xi_t-1}{\xi_t}\ln\frac{m_Q^2}{m_t^2} -
G(\xi_t)\right) N_R^2N_L^2
F(\xi_t)^{-1}\label{delta_penguin}
\end{eqnarray}
Here $\delta_c$ accounts for the $c$--quark contribution
(\ref{L_sdnunu_SM}). Substituting the numerical values of the
parameters of New Physics from (\ref{numbers}) and the numerical
value $F(\xi_t) = 2.744$ (\ref{Fxi}) and neglecting for the moment
$\delta_c$ we obtain:
\begin{equation}
\delta_p = 0.04. \label{delta_penguin_n}
\end{equation}

The correction $\delta_p$ describes the deviation of the
$t$--quark dominated penguin amplitudes of the decays $K_L
\rightarrow \pi^0 \nu \bar{\nu}$, $B \rightarrow \pi \nu
\bar{\nu}$ and $B \rightarrow K \nu \bar{\nu}$ from the SM. The
decay $K_L \rightarrow \pi^0 \nu \bar{\nu}$ proceeds only via the
CP violation mechanism and includes the CKM factor ${\rm
Im}\left(V_{is}V*_{id}\right)$ which is approximately the same for
both the $c$-- and $t$--quarks inside the loop. Thus the
$c$--quark contribution is damped by a factor
$\dfrac{m_c^2}{m_t^2} \approx 10^{-4}$ as compared to $t$--quark.
In the case of the B-meson decays the CKM factors are of the same
order for the $c$-- and $t$--quarks inside the loop and the charm
contribution is damped again by the factor $\dfrac{m_c^2}{m_t^2}$.
For these decays $\delta_c$ is negligible.

The case of the $K^+ \rightarrow \pi^+ \nu \bar{\nu}$ is more
involved. When calculating its probability we have to take into
account the contribution of the lepton box diagram with the
$c$--quark inside the loop. This effect turns out to be quite
significant \cite{buras}, $\tilde{\delta}_c \approx
0.3\times\tilde{\delta}_p$. As a result, for the $K^+ \rightarrow
\pi^+ \nu \bar{\nu}$ decay we obtain $\tilde{\delta}_p \approx
0.03$ and the corresponding branching ratio increases by
approximately $\tilde{\delta}_{br} = 2\tilde{\delta}_p$ as
compared to the SM:
\begin{equation}
\tilde{\delta}_{br} = 0.06. \label{tilde_delta_br}
\end{equation}

The increase of the branching ratios of the $K_L \rightarrow \pi^0
\nu \bar{\nu}$, $B \rightarrow \pi \nu \bar{\nu}$ and $B
\rightarrow K \nu \bar{\nu}$ decays will be
\begin{equation}
\delta_{br} = 2\delta_p = 0.08. \label{delta_br}
\end{equation} Thus, for $5\rm\ TeV$ mass of the isosinglet quark $Q$ the
branching ratios of the $Z$--penguin originated decays $K_L
\rightarrow \pi^0 \nu \bar{\nu}$, $B \rightarrow \pi \nu
\bar{\nu}$ and $B \rightarrow K \nu \bar{\nu}$ are $8\%$ larger
than in the SM.

To obtain the constraints on the $Q$ mass\footnote{From now on,
the parameter $M$ will no longer be a constant. To keep $\mu_L$
equal to its value in equation (\ref{numbers}) and, which is more
important, $m_t$ equal to its experimental value we adjust $\mu_R$
according to (\ref{masses}).} from the experimental data we will
rewrite $\delta_p$ (\ref{delta_penguin}) in a more convenient
form:
\begin{eqnarray}
\delta_p &\approx& \frac{\mu_L^2}{M^2} \left[ -2\left(
1-\frac{m_t^2}{\mu_L^2} \right)^2 + \frac{\mu_L^2}{m_t^2F(\xi_t)}
\right.
 + \left. 2\frac{(\xi_t-1)/\xi_t\cdot\ln(M^2/m_t^2)-G(\xi_t)}{F(\xi_t)} \right]\nonumber\\
 &\equiv& \frac{\mu_L^2}{M^2}\ f(\mu_L,\ \ln M).\label{delta_penguin_prop}
\end{eqnarray} Since the dependence of $f(\mu_L,\ \ln M)$ on $M$
 is only logarithmical and thus
very weak, we will use its value at $M = 5\rm\ TeV$:
\begin{equation}
f(\mu_L,\ \ln 5\rm\ TeV) \approx 5.0.\label{fmu_lnM}
\end{equation}

In this way we obtain a simple expression for the branching ratios
in our model:

\begin{eqnarray}
&&Br(K,\ B \rightarrow \pi, K \nu \bar{\nu})_{NP}=Br(K,\ B
\rightarrow \pi, K \nu \bar{\nu})_{SM} \cdot\left(1+2\ f(\mu_L,\ln
M)\ \frac{\mu_L^2}{M^2}\right),
\label{delta_max_def}\end{eqnarray} which can be straightforwardly
compared with the experimental data (for the $K^+ \rightarrow
\pi^+ \nu \bar{\nu}$ decay one should substitute $0.8$ instead of
unity in parentheses in (\ref{delta_max_def})).

The present-day results for the branching ratios are the following
\cite{pdg}:
\begin{eqnarray}
&&Br(K^+ \rightarrow \pi^+ \nu \bar{\nu}) =
(1.5^{+1.3}_{-0.9})\times 10^{-10},\nonumber\\
&&Br(K_L \rightarrow \pi^0 \nu \bar{\nu}) < 5.9 \times 10^{-7},\nonumber\\
&&Br(B_u \rightarrow \pi^+ \nu \bar{\nu}) < 1.0 \times 10^{-4},\nonumber\\
&&Br(B_u \rightarrow K^+ \nu \bar{\nu}) < 5.2 \times 10^{-5}.
\label{br_num_1}\end{eqnarray}

New data were obtained in August 2007 \cite{hfag}:
\begin{equation}Br(B_u \rightarrow K^+ \nu
\bar{\nu}) < 1.4 \times 10^{-5}. \label{br_num_3}\end{equation}
and in December 2007 \cite{kpinunuahn}:
\begin{equation}Br(K_L \rightarrow \pi^0 \nu \bar{\nu}) < 6.7 \times 10^{-8}.
\label{br_num_2}\end{equation}

The current state of the experiment does not allow us to make any
conclusions concerning the existence of the New Physics. The
future plans are the following:
\begin{itemize}
\item the measurement of $Br(K^+ \rightarrow \pi^+ \nu \bar{\nu})$
at the CERN SPS NA62 experiment with $\approx 10\%$ accuracy, the
data taking is planned for 2009--2010 \cite{spsna62}; \item the
measurement of $Br(B_u \rightarrow K^+ \nu \bar{\nu})$ at the
Super B Factory experiment with the accuracy $\leq 20\%$ by
2014--2015 \cite{superB}; \item the measurement of $Br(K^+
\rightarrow \pi^+ \nu \bar{\nu})$ at the J-PARC experiment with
the accuracy $\leq 20\%$ after 2012--2013 \cite{jparc1}; \item the
measurement of $Br(K_L \rightarrow \pi^0 \nu \bar{\nu})$ at the
J-PARC experiment with the accuracy $\leq 10\%$ after 2010
\cite{jparc2}.
\end{itemize}



\section{Conclusions.}
The branching ratios $Br(K \rightarrow \pi \nu \bar{\nu})$, $Br(B
\rightarrow \pi \nu \bar{\nu})$ and $Br(B \rightarrow K \nu
\bar{\nu})$ get in our model up to $10\%$ corrections for the mass
of the isosinglet quark $M=5\rm\ TeV$. The uncertainties coming
from the poor knowledge of the CKM matrix elements should
considerably improve in the near future, while the experimental
data on the probabilities of the rare decays analyzed in the paper
should also appear in due time (\cite{spsna62}, \cite{superB},
\cite{jparc1} and \cite{jparc2}). The proper accuracy of the data
will allow to discover New Physics or to establish the lower
bounds on the mass of the heavy quark $Q$.



\section{Acknowledgments.}
We thank A. E. Bondar for emphasizing the importance of the
consideration of $K \rightarrow \pi \nu \bar{\nu}$ decays in our
model and for the helpful information on the experimental
prospects concerning $B \rightarrow K \nu \bar{\nu}$ decays. Our
research is supported by the grants NSh-4568.2008.2,
RFBR-08-02-00494 and partially by Rosatom. The research of P. N.
K. is also supported by the grants NSh-8004.2006.2,
RFBR-07-02-00878 and by the Dmitry Zimin Dynasty Foundation. The
research of M. I. V. is also supported by the grant
RFBR-07-02-00021.

\end{document}